\documentclass[twocolumn]{revtex4}

\usepackage{amsmath,amssymb}
\usepackage{graphicx}% Include figure files
\usepackage{color}% Include colors for document elements
\usepackage{dcolumn}% Align table columns on decimal point
\usepackage{bm}% bold math

\begin{document}
\title{Quantum Monte Carlo for large chemical systems: Implementing efficient
strategies for petascale platforms and beyond}

\author{Anthony Scemama, Michel Caffarel}
\affiliation{Laboratoire de Chimie et Physique Quantiques, CNRS-IRSAMC,
Universit\'e de Toulouse, France.}
\author{Emmanuel Oseret, William Jalby}
\affiliation{Exascale Computing Research Laboratory, GENCI-CEA-INTEL-UVSQ,
Universit\'e de Versailles Saint-Quentin, France.}

\begin{abstract}
Various strategies to implement efficiently QMC simulations for large chemical
systems are presented. These include: i.) the introduction of an efficient
algorithm to calculate the computationally expensive Slater matrices. This
novel scheme is based on the use of the highly localized character of {\it
atomic} Gaussian basis functions (not the {\it molecular} orbitals as usually
done), ii.) the possibility of keeping the memory footprint minimal, iii.) the
important enhancement of single-core performance when efficient optimization
tools are employed, and iv.) the definition of a universal, dynamic,
fault-tolerant, and load-balanced framework adapted to all kinds
of computational platforms (massively parallel machines, clusters, or
distributed grids). These strategies have been implemented in the QMC=Chem code
developed at Toulouse and illustrated with numerical applications on small
peptides of increasing sizes (158, 434, 1056 and 1731 electrons).  Using
10k--80k computing cores of the Curie machine (GENCI-TGCC-CEA, France)
QMC=Chem has been shown to be capable of running at the petascale level, thus
demonstrating that for this machine a large part of the peak performance can be
achieved. Implementation of large-scale QMC simulations for future exascale
platforms with a comparable level of efficiency is expected to be feasible.
\end{abstract}

\maketitle

%*****************Graphical Table of Contents******************** THIS IS MANDATORY *******************

% makes references listed with 1., 2., etc.  
  \makeatletter
  \renewcommand\@biblabel[1]{#1.}
  \makeatother

\bibliographystyle{apsrev}

\clearpage

\section{Introduction}
Quantum Monte Carlo (QMC) is a generic name for a large class of stochastic approaches solving the Schr\"odinger equation by using random walks.
In the last forty years they have been extensively used in several fields of physics including
nuclear physics,\cite{qmc-nuclear} condensed-matter physics,\cite{rmp-foulkes} spin systems,\cite{suzuki}, quantum liquids,\cite{rmp-ceperley} infrared 
spectroscopy,\cite{qmc-vib1,qmc-vib2} etc.
In these domains QMC methods are usually considered as routine methods and even in most cases as state-of-the-art approaches. 
In sharp contrast, this is not yet the case for the electronic structure problem of quantum chemistry where QMC\cite{lester,qmc} 
is still of confidential use 
when compared to the two well-established methods of the domain (Density Functional Theory (DFT) and post-Hartree-Fock methods). 
Without entering into the details of the forces and weaknesses of each approach, a major limiting aspect of QMC hindering 
its diffusion is the high computational cost of the simulations for realistic systems.

However --- and this is the major concern of this work --- a unique and
fundamental property of QMC methods is their remarkable adaptation to High
Performance Computing (HPC) and, particularly, to massively parallel
computations. 
In short, the algorithms are simple and repetitive, central memory requirements may be kept limited whatever the system size, and
I/O flows are negligible. As most Monte Carlo algorithms, the computational effort is almost exclusively concentrated on pure CPU 
(``number crunching method'') and the execution time is directly proportional to the number of Monte Carlo steps performed.
In addition, and this is a central point for massive parallelism,  calculations of averages can be decomposed at will: 
$n$ Monte Carlo steps over a single processor being equivalent to $n/p$ Monte Carlo steps over $p$ processors 
with no communication between the processors (apart from the initial/final data transfers). 
Once the QMC algorithm is suitably implemented the maximum gain of parallelism (ideal scalability) should be expected.

A most important point is that mainstream high-level quantum chemistry methods do not enjoy such a remarkable property. 
Essentially based on iterative schemes defined with the framework of linear algebra they involve the manipulation and 
storage of extremely large matrices and their adaptation to extreme parallelism and low-memory-footprint implementation 
is intrinsically problematic. 

Now, in view of the formidable development of computational platforms, particularly in terms of number of computing cores (presently up to a few hundreds 
of thousands and many more to come) the practical bottleneck associated with the high computational cost of QMC 
is expected to become much less critical and thus QMC may become in the coming years a method of practical use for treating chemical problems 
out of the reach of present-day approaches. 
Following this line of thought a number of QMC groups are presently working on implementing strategies allowing their QMC codes to run 
efficiently on very large-scale parallel computers.\cite{esler,kim,gillan}
Essentially, most strategies rely on massive parallelism and on some efficient treatment (``linear-scaling''-type algorithms) for dealing with 
the matrix computations and manipulations that represent the most CPU-expensive part of the algorithm.

Here, we present several strategies implemented in the QMC=Chem code developed
in our group at the University of Toulouse.\cite{qmcchem} A number of actual
simulations realized on the Curie machine at the French GENCI-TGCC-CEA
computing center with almost ideal parallel efficiency in the range 
10 000--80 000 cores and reaching the petascale level have been realized.

The contents of this paper is as follows. In The first section, a brief account
of the QMC method employed is presented. Only those aspects essential to the
understanding of the computational aspects discussed in this work are given.
In second section, the problem of computing efficiently the Slater matrices at
the heart of the QMC algorithm (computational hot spot) is addressed.  A novel
scheme taking advantage of the highly-localized character of the {\it atomic}
Gaussian basis functions (not the {\it molecular} orbitals as usually done) is
proposed.  A crucial point is that the approach is valid for an arbitrary
molecular shape ({\it e.g.} compact molecules), there is no need of considering
extended or quasi-one-dimensional molecular systems as in linear-scaling
approaches.  The third section discusses the overall performance of the code
and illustrates how much optimizing the single-core performance of the specific
processor at hand can be advantageous. The fourth section is devoted to the way
our massively parallel simulations are deployed on a general computational
platform and, particularly, how fault-tolerance is implemented, a crucial
property for any large-scale simulation.  Finally, a summary of the various
strategies proposed in this work is presented in the last section.

\section{The quantum Monte Carlo method}
\label{QMC}
In this work we shall consider a variant of the Fixed-Node Diffusion Monte Carlo (FN-DMC) approach, the standard quantum Monte Carlo method used
in computational chemistry. Here, we shall insist only on the aspects needed for understanding the rest of the work.
For a complete presentation of the FN-DMC method the reader is referred, {\it e.g} to \cite{rmp-foulkes},\cite{lester}, or \cite{qmc} and references therein.

\subsection{Fixed-node diffusion Monte Carlo (FN-DMC)} 
{\it Diffusion Monte Carlo}. In a diffusion Monte Carlo scheme, a finite population of ``configurations'' or ``walkers'' moving in 
the $3N$-dimensional space ($N$, number of electrons) is introduced. A walker is described by a $3N$-dimensional vector
${\bf R}\equiv ({\bf r}_1,{\bf r}_2,...,{\bf r}_N)$ giving the positions of the $N$ electrons.  
At each Monte Carlo step each walker of the population is diffused and drifted according to
\begin{equation}
{\bf R^\prime}= {\bf R}+\tau {\bf b}({\bf R}) + \sqrt{\tau} {\bf \eta}
\label{proba}
\end{equation}
where $\tau$ is a small time-step, ${\bf \eta}$ a Gaussian vector ($3N$
independent normally distributed components simulating a free Brownian
diffusion), and ${\bf b}({\bf R})$ the drift vector given by
\begin{equation}
{\bf b}({\bf R}) \equiv \frac{{\bf \nabla} \psi_T({\bf R})}{\psi_T({\bf R})},
\label{drift}
\end{equation}
where $\psi_T$, the trial wave function, is a known computable approximation of the exact wavefunction.
At the end of this drift/diffusion step each walker is killed, kept unchanged, 
or duplicated a certain number of times proportionally to the branching weight $w$ given by
\begin{equation}
w=e^{-\frac{\tau}{2} [(E_L({\bf R^\prime})-E_T) + (E_L({\bf R})-E_T)]}
\label{w}
\end{equation} 
where $E_T$ is some reference energy and $E_L$ the local energy defined as
\begin{equation}
E_L({\bf R}) \equiv \frac{ H\psi_T({\bf R})}{\psi_T({\bf R})}.
\label{localenergy}
\end{equation}
The population is propagated and after some equilibrium time it enters a stationary regime 
where averages are evaluated. As an important example, the exact energy may be obtained as 
the average of the local energy.\\
\\
\noindent
{\it The Fixed-Node approximation}. Apart from the statistical and the short-time (finite time-step) errors which can be made arbitrary small, 
the only systematic error left in a DMC simulation is the so-called Fixed-Node (FN) error. This error results from the fact that the nodes 
of the trial wavefunction [defined as the $(3N-1)$-dimensional
hyper-surface where $\Psi_T({\bf R})=0$] act as infinitely repulsive barriers for the walkers [divergence of the drift vector, Eq.(\ref{drift})].
Each walker is thus trapped forever within the nodal pocket delimited by the nodes of $\Psi_T$ where it starts from. 
When the nodes of $\psi_T$ coincide with the exact nodes, the algorithm is exact. If not,
a variational fixed-node error is introduced. However, with the standard trial wavefunctions used, 
this error is in general small,\cite{notefn} a few percent of the correlation energy for total energies.

\subsection{Parallelizing FN-DMC} 
Each Monte Carlo step is carried out {\it independently} for each walker of the
population. The algorithm can thus be easily parallelized over an arbitrary
number of processors by distributing the walkers among the processors, but
doing this implies synchronizations of the CPUs since the branching step
requires that all the walkers have first finished their drifted-diffusion step.

To avoid this aspect, we have chosen to let each CPU core manage its own
population of walkers without any communication between the populations.
On each computing unit a population of walkers is propagated and the various
averages of interest are evaluated. At the end of the simulation, the averages
obtained on each processor are collected and summed up to give the final
answers. Regarding parallelism the situation is thus ideal since, apart from
the negligible initial/final data transfers, there are no communications among
processors.

The only practical problem left with FN-DMC is that the branching
process causes fluctuations in the population size and thus may lead to
load-balancing problem among processors. More precisely, nothing prevents the
population size from decreasing or increasing indefinitely during the Monte
Carlo iterations. To escape from this, a common solution consists in forcing
the number of walkers not to deviate too much from some target value for the
population size by introducing a population control step.  It is usually
realized by monitoring in time the value of the reference energy $E_T$ via a
feedback mechanism, see {\it e.g.} \cite{umrigar1993}.  The price to pay is the
introduction of some transient load imbalances and inter-processor
communications/synchronization to redistribute walkers among computing cores,
inevitably degrading the parallel speed-up. This solution has been adapted by
several groups and some tricks have been proposed to keep this problem under
control.\cite{esler,kim,gillan,pop_ceperley}

Here, we propose to avoid this problem directly from the beginning by employing
a variant of the FN-DMC working with a constant number of walkers.  Several
proposals can be found in the literature, e.g. \cite{sorella,srmc}. Here, we
shall employ the method described in Ref.\cite{srmc}. In this approach the
branching step of standard DMC is replaced by a so-called reconfiguration step.
Defining the {\it normalized} branching weights as follows
\begin{equation}
p_k= \frac{w_k}{\sum_{i=1}^M w_i}
\end{equation}
the population of walkers is ``reconfigured'' by drawing at each step $M$
walkers among the $M$ walkers according to the probabilities $p_k$.  At
infinite population, this step reduces to the standard branching step where
walkers are deleted or duplicated proportionally to the weight $w$.  At finite
$M$ the normalization factor $\sum_{i=1}^M w_i$ is no longer constant and a
finite-population bias is introduced.  To remove this error and recover the
exact averages, a global weight given as the product of the total population
weights for all preceding generations must be included into the averages.  Note
that this algorithm enables the possibility to use small walker populations on
each core since there is no finite-population bias (typically, we use 10 to 100
walkers per core).  For all details the reader is referred to \cite{srmc}.

\subsection{Critical CPU part} 
\label{cpu}
At each Monte Carlo step the CPU effort is almost completely dominated by the evaluation of the wavefunction $\Psi_T$
and its first and second derivatives (computational hot spot). More precisely, for each walker the values of the trial wavefunction, $\Psi_T$, its
first derivatives with respect to all $3N$-coordinates [drift vector, Eq.(\ref{drift})], and its Laplacian $\nabla^2 \Psi_T$ [kinetic part of the 
local energy, Eq.(\ref{localenergy})] are to be calculated. It is essential that such calculations be as efficient as possible since in realistic applications 
their number may be very large (typically of the order of $10^{9}-10^{12}$).

A common form for the trial wavefunction is
\begin{widetext}
\begin{equation}
\Psi_T({\bf R}) = e^{J({\bf R})} \sum_{K=(K_\uparrow,K_\downarrow)} c_K {\rm Det}_{K_\uparrow}({\bf r}_1,...,{\bf r}_{N_\uparrow}) {\rm Det}_{K_\downarrow}({\bf r}_{N_\uparrow+1},...,{\bf r}_N).
\label{psit}
\end{equation}
\end{widetext}
where the electron coordinates of the $N_\uparrow$ (resp. $N_\downarrow$) electrons of spin $\uparrow$ (resp. $\downarrow$) have been distinguished,
$ N= N_\uparrow+N_\downarrow$.
In this formula $e^{J({\bf R})}$ is the Jastrow factor describing explicitly the electron-electron
interactions at different levels of approximations. A quite general form may be written as
\begin{eqnarray}
J({\bf R}) &=& \sum_\alpha  U^{(e-n)}(r_{i \alpha}) + \sum_{i,j}  U^{(e-e)}(r_{ij})  \\\nonumber
&&+ \sum_{\alpha i,j}  U^{(e-e-n)}(r_{ij},r_{i \alpha},r_{j \alpha}) +...
\end{eqnarray}
where $r_{ij}= |{\bf r}_i-{\bf r}_j|$ is the inter-electronic distance and
$r_{i\alpha}= |{\bf r}_i-{\bf Q}^{\alpha}|$, the distance between electron $i$
and nucleus $\alpha$ located at ${\bf Q}^{\alpha}$.  Here $U$'s are simple
functions and various expressions have been employed in the literature. 
The Jastrow factor being essentially local, short-ranged expressions can
be employed and the calculation of this term is usually a small contribution to
the total computational cost. As a consequence, we shall not discuss further
the computational aspect of this term here.

The second part of the wavefunction describes the shell-structure in terms of
single-electron molecular orbitals and is written as a linear combination of
products of two Slater determinants, one for the $\uparrow$ electrons and the
other for the $\downarrow$ electrons. Each Slater matrix is built from a set of
molecular orbitals $\phi_i({\bf r})$ usually obtained from a preliminary DFT or
SCF calculations. The $N_{\rm orb}$ molecular orbitals (MOs) are expressed as a
sum over a finite set of $N_{\rm basis}$ basis functions (atomic orbitals, AOs)
\begin{equation}
\phi_i({\bf r})= \sum_{j=1}^{N_{\rm basis}} a_{ij} \chi_j({\bf r})
\label{mo}
\end{equation}
where the basis functions $\chi_j({\bf r})$ are usually expressed as a product
of a polynomial and a linear combination of Gaussian functions. In the present
work the following standard form is employed
\begin{equation}
\chi({\bf r})= (x-Q_x)^{n_x} (y-Q_y)^{n_y} (z-Q_z)^{n_z}
g({\bf r})
\end{equation}
with
\begin{equation}
g({\bf r})= 
\sum_k c_{k} e^{-\gamma_k ({\bf r} - {\bf Q})^2}.
\label{Gaussian}
\end{equation}
Here ${\bf Q}=(Q_x,Q_y,Q_z)$ is the vector position of the nucleus-center of
the basis function, ${\bf n}=(n_x,n_y,n_z)$ a triplet of positive integers,
$g({\bf r})$ is the spherical Gaussian component of the AO, and $\gamma_k$ its
exponents. The determinants corresponding to spin $\uparrow$-electrons are
expressed as 
\begin{equation}
{\rm Det}_K{_\uparrow}({\bf r}_1,...,{\bf r}_{N_\uparrow}) 
= {\rm Det}
\left(
\begin{array}{ccc}
\phi_{i_1}({\bf r}_1) & \ldots & \phi_{i_1}({\bf r}_{N_{\uparrow}})\\
\vdots                           & \vdots & \vdots \\
\phi_{i_{N_{\uparrow}}}({\bf r}_1) & \ldots & \phi_{i_N{_{\uparrow}}}
({\bf r}_{N_{\uparrow}})\\
\end{array}
\right)
\end{equation}
where $K{_\uparrow}$ is a compact notation for denoting the set of indices
$\{i_1,...,i_{N_\uparrow}\}$ specifying the subset of the molecular orbitals
used for this particular Slater matrix. A similar expression is written for
spin $\downarrow$-electrons.

In contrast to the calculation of the Jastrow factor, the evaluation of the
determinantal part of the wavefunction and its derivatives is critical. To
perform such calculations we employ a standard approach\cite{lester} consisting
in calculating the matrices of the first and second (diagonal) derivatives of
each molecular orbital $\phi_i$ with respect to the three space variables
$l=x,y,z$ evaluated for each electron position ${\bf r}_j$, namely
\begin{equation}
D_{l,ij}^{(1)} \equiv \frac{\partial \phi_i({\bf r}_j)}{\partial x^j_l}
\end{equation}
\begin{equation}
D_{l,ij}^{(2)} \equiv \frac{\partial^2 \phi_i({\bf r}_j)}{\partial {x^j_l}^2}
\end{equation}
and then computing the inverse $D^{-1}$ of the Slater matrix defined as $D_{ij}=\phi_i({\bf r}_j)$. 
The drift components and the Laplacian corresponding to the determinantal part of the trial wavefunction are thus evaluated as simple vector-products
\begin{equation}
\frac{1}{{\rm Det}({\bf R})} \frac{\partial {\rm Det}({\bf R})}{\partial x^i_l}= \sum_{j=1,N} D_{l,ij}^{(1)} D_{ji}^{-1}
\end{equation}
\begin{equation}
\frac{1}{{\rm Det}({\bf R})} \frac{\partial^2 {\rm Det}({\bf R})}{\partial {x^i_l}^2}= \sum_{j=1,N} D_{l,ij}^{(2)} D_{ji}^{-1}
\end{equation}

From a numerical point of view, the computational time $T$ needed to evaluate such quantities as a function of the number of electrons $N$ 
scales as ${\cal O}(N^3)$
\begin{equation}
T= \alpha N^3 + \beta N^3.
\end{equation}
The first $N^3$-term results from the fact that the $N^2$ matrix elements of the Slater matrices are to be computed, each element being expressed
in terms of the $N_{\rm basis} \sim N$ basis functions needed to reproduce an arbitrary delocalized molecular orbital.
The second $N^3$-term is associated with the generic cubic scaling of any linear algebra method for inverting a general matrix. 

\section{Exploiting the highly localized character of atomic basis functions}
\label{gaussiansec}
As seen in the previous section, one of the two computational hot spots of QMC is the 
calculation of the derivatives of the determinantal part of the trial wave
function for each electronic configuration $({\bf r}_1,...,{\bf r}_{N})$ at each Monte Carlo step. To be more precise, the $N_{\rm orb}$ molecular
orbitals (MO) used in the determinantal expansion (\ref{psit}) are to be
computed (here, their values will be denoted as ${\bf C}_1$) together with their
first derivatives with respect to $x$, $y$, and $z$ (denoted ${\bf C}_2,{\bf C}_3,{\bf C}_4$) and
their Laplacians (denoted ${\bf C}_5$). Calculations are made in single precision
using an efficient matrix product routine we describe now.  The matrix
products involve the matrix of the MO coefficients $a_{ij}$,
Eq.(\ref{mo}) (here denoted as ${\bf A}$) the matrix of the atomic Gaussian basis functions evaluated at all electronic positions,
$\chi_j({\bf r}_i)$ (denoted ${\bf B}_1$), their first
derivatives (denoted ${\bf B}_2,{\bf B}_3,{\bf B}_4$) and Laplacians (denoted ${\bf B}_5$).
The five matrix products are written under the convenient form
\begin{equation}
 {\bf C}_i = {\bf A} {\bf B}_i \;\;\;\;i=1,5
\label{abc}
\end{equation}
Note that matrix ${\bf A}$ remains constant during the simulation while matrices ${\bf B}_i$ and
${\bf C}_i$ depend on electronic configurations. The matrix sizes are as
follows: $N_{\rm orb} \times N$ for the ${\bf C}_i$'s, $N_{\rm orb} \times
N_{\rm basis}$ for ${\bf A}$, and $N_{\rm basis} \times N$ for ${\bf B}$. In
practical applications $N_{\rm orb}$ is of the order of $N$ while
$N_{\rm basis}$ is greater than $N$ by a factor 2 or 3 for standard
calculations and much more when using high-quality larger basis sets. The
expensive part is essentially dominated by the $N_{\rm basis}$ multiplications.
The total computational effort is thus of order $N_{\rm orb} \times N
\times N_{\rm basis}$, {\it i.e.} $\sim {\cal O}(N^3)$.

The standard approach proposed in the literature for reducing the $N^3$-price
is to resort to the so-called linear-scaling or ${\cal
O}(N)$-techniques.\cite{ls1,ls2,ls3,ls4,ls5,ls6}
The basic idea consists in introducing {\em spatially localized} molecular
orbitals instead of the standard delocalized (canonical) ones obtained from
diagonalization of reference Hamiltonians (usually, Hartree-Fock or Kohn-Sham).
Since localized orbitals take their value in a finite region of space ---
usually in the vicinity of a fragment of the molecule --- the number of basis
set functions $N_{\rm basis}$ needed to represent them with sufficient accuracy
becomes essentially independent of the system size (not scaling with $N$ as in the case of canonical ones).
In addition to this, each electron contributes only to a small subset of the
localized orbitals (those non-vanishing in the region where the electron is
located).  As a consequence, the number of non-vanishing matrix elements of the
${\bf C}_i$ matrices no longer scales as $N_{\rm orb} \times N \sim N^2$ but linearly with $N$. Furthermore, each matrix element whose
computation was proportional to the number of basis set used, $N_{\rm basis}
\sim N$, is now calculated in a finite time independent of the system
size. Putting together these two results, we are led to a linear dependence 
of the computation of the ${\bf C}_i$ matrices upon the number of electrons.

Here, we choose to follow a different path. Instead of localizing the canonical
molecular orbitals we propose to take advantage of the localized
character of the underlying {\em atomic} Gaussian basis set functions.  The
advantages are essentially three-fold: 
\begin{enumerate}
\item the atomic basis set functions are naturally localized {\em independently
of the shape of the molecule}. This is the most important point since the
localization procedures are known to be effective for chemical systems having a
molecular shape made of well-separated sub-units (for example, linear systems)
but much less for general compact molecular systems that are ubiquitous in
chemistry.
\item the degree of localization of the standard atomic Gaussian functions is
much larger than that obtained for molecular orbitals after localization (see
results below) 
\item by using the product form, Eq.(\ref{abc}), the localized nature of the
atomic Gaussian functions can be exploited very efficiently (see next section).
\end{enumerate}
In practice, when the value of the spherical Gaussian part $g({\bf r})$ of an
atomic orbital function $\chi({\bf r})$ is smaller than a given threshold
$\epsilon=10^{-8}$, the value of the AO, its gradients and Laplacian are
considered null.
This property is used to consider the matrices ${\bf B}_1,\dots,{\bf B}_5$ as sparse.
However, in contrast with linear-scaling approaches, the MO matrix ${\bf A}$ is not
considered here as sparse. We shall come back to this point later.
To accelerate the calculations, an atomic radius is computed as the distance
beyond which all the Gaussian components $g({\bf r})$ of the atomic orbitals
$\chi{(\bf r)}$ centered on the nucleus are less than $\epsilon$. If an electron
is farther than the atomic radius, all the AO values, gradients and Laplacians
centered on the nucleus are set to zero.

The practical implementation to perform the matrix products is as
follows.  For each electron, the list of indices (array {\tt indices} in
what follows) where $g({\bf r})> 0$ is calculated. Then, the practical
algorithm can be written as
\begin{verbatim}
C1 = 0. 
C2 = 0. 
C3 = 0.
C4 = 0. 
C5 = 0.
do i=1, Number of electrons
 do k=1, Number of non-zero AOs for electron i
  do j=1, Number of molecular orbitals
   C1(j,i) += A(j,indices(k,i))*B1(k,i)
   C2(j,i) += A(j,indices(k,i))*B2(k,i)
   C3(j,i) += A(j,indices(k,i))*B3(k,i)
   C4(j,i) += A(j,indices(k,i))*B4(k,i)
   C5(j,i) += A(j,indices(k,i))*B5(k,i)
  end do
 end do
end do
\end{verbatim}
(where \verb$x += y$ denotes \verb$x = x + y$).

This implementation allows to take account of the sparsity of the ${\bf B}$ matrices,
while keeping the efficiency due to a possible vectorization of the inner loop.
The load/store ratio is 6/5 (6 load-from-memory instructions, 5 store-to-memory instructions) 
in the inner loop~: the elements of ${\bf B}_n$ are constant
in the inner loop (in registers), and the same element of ${\bf A}$ is used at each
line of the inner loop (loaded once per loop cycle). As store operations are
more expensive than load operations, increasing the load/store ratio improves
performance as will be shown in the next section.
Using this algorithm, the scaling of the matrix products is expected to
drop from ${\cal O}(N^3)$ to a scaling roughly equal to ${\cal O}(N^2)$ (in a
regime where $N$ is large enough, see discussion in the next section). 
Let us now illustrate such a property in the applications to follow.

\begin{figure}
\begin{center}
\includegraphics[width=\columnwidth,keepaspectratio=true]{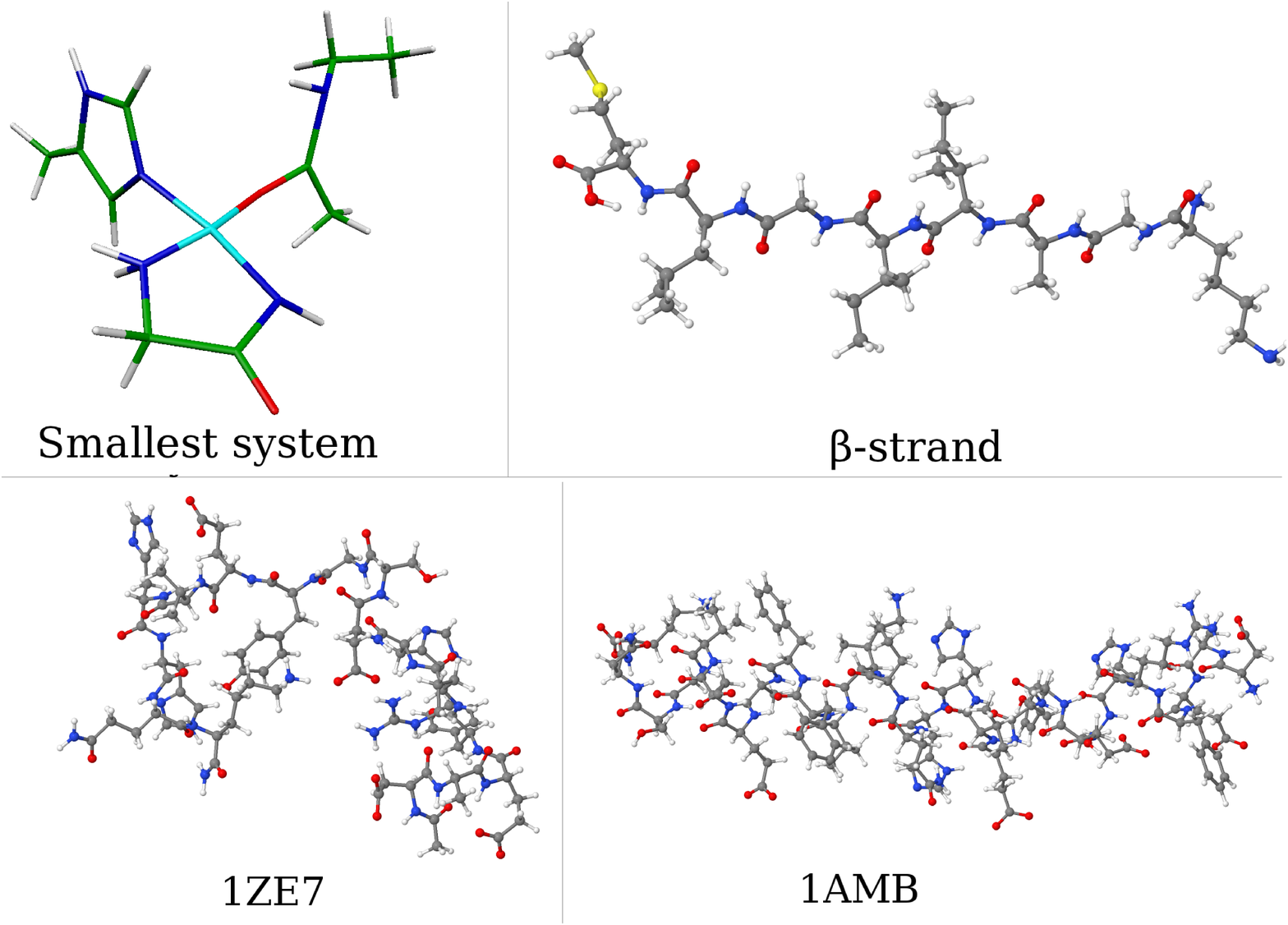}
\end{center}
\caption{Molecular systems used as benchmarks.}
\label{fig1}
\end{figure}

The different systems used here as benchmarks are represented in Figure
\ref{fig1}. The trial wavefunctions used for describing each system are
standard Hartree-Fock wavefunctions (no Jastrow factor) with molecular orbitals
expressed using various Gaussian basis sets. System~1 is a copper complex with
four ligands having 158 electrons and described with a cc-pVDZ basis set.
System~2 is a polypeptide taken from reference\cite{alzheimer} (434 electrons
and 6-31G$^*$ basis set).
System~3 is identical to System~2 but using a larger basis set, namely the
cc-pVTZ basis set. System~4 is the 1ZE7 molecule from the Protein Data Bank
(1056 electrons, 6-31G$^*$), and System~5 is the 1AMB molecule from the Protein
Data Bank (1731 electrons, 6-31G$^*$).

Table~\ref{tab1} shows the level of sparsity of the matrices ${\bf A}$ ($A_{ij}
\equiv a_{ij}$) and ${\bf B}_1$ (${B_1}_{ij} \equiv \chi_i({\bf r}_j)$) for the five
systems (matrices ${\bf B}_n$ with $n>1$ behave as ${\bf B}_1$ with respect to sparsity). 
As seen the number of basis set functions employed is proportional to the
number of electrons with a factor ranging from about 2.2 to 6.8.

Regarding the matrix ${\bf A}$ of MO coefficients the results are given both
for standard canonical (delocalized) MOs and for localized orbitals. 
To get the latter ones, different localization schemes have been
applied.\cite{boys,pipek,aquilante} However, they essentially lead to
similar results. Here, the results presented are those obtained by using the
Cholesky decomposition of the density matrix expressed in the AO basis
set.\cite{aquilante}
As seen the level of sparsity of the matrix ${\bf A}$ is low. Although it increases
here with the system size it remains modest for the largest size (there are
still about one third of non-zero elements). Of course, such a result strongly
depends on the type of molecular system considered (compact or not compact) and
on the diffuse character of the atomic basis set. Here, we have considered typical systems of biochemistry.

Next, the level of sparsity of the ${\bf B}$ matrices is illustrated. The percentage of
non-zero values of $\chi_i({\bf r}_j)$ has been obtained as an average over a
Variational Monte Carlo run. In sharp contrast with MOs the atomic orbitals are
much more localized, thus leading to a high level of sparsity. For the largest
system, only 3.9$\%$ of the basis function values are non-negligible.

In the last column of the table the maximum number of non-zero elements
obtained for all columns of the matrix (i.e, a value of electron position)
during the entire Monte Carlo simulation is given. A first remark is that this
number is roughly constant for all system sizes. A second remark is that the
percentage of non-zero values is only slightly greater than the average, thus
showing that the ${\bf B}$ matrices can be considered sparse during the whole
simulation, not only in average. As an important consequence, the loop over the
number of non-zero AOs for each electron in the practical algorithm presented
above (loop over $k$ index) is expected to be roughly constant as a function of
the size at each Monte Carlo step. This latter remark implies for this part an
expected behavior of order ${\cal O}(N^2)$ for large $N$. 
Let us now have a closer look at the actual performance of the code.

\section{Overall performance of QMC=Chem}
\label{performance}

When discussing performance several aspects must be considered. A first one,
which is traditionally discussed, is the formal scaling of the code as a
function of the system size $N$ ($N \sim$ number of electrons). As already
noted, due to the innermost calculation, products, and inversion of
matrices, such a scaling is expected to be cubic, ${\cal O}(N^3)$. However,
there is a second important aspect, generally not discussed, which is related
to the way the expensive innermost floating-point operations are implemented
and on how far and how efficiently the potential performance of the processor
at hand is exploited. In what follows we shall refer to this aspect as
``single-core optimization''.  It is important to emphasize that such an aspect
is by no way minor and independent on the previous ``mathematical'' one.

To explicit this point, let us first recall that the computational time $T$
results essentially from two independent parts, the first one resulting from
the computation of the matrix elements, $T_1 \sim \alpha N^3$ and the second
one from the inversion of the Slater matrix, $T_2 \sim \beta N^3$. 
Now, let us imagine that we have been capable of devising a highly efficient
linear-scaling algorithm for the first contribution such that $T_1 \sim
\epsilon N <<  T_2 $ within the whole range of system sizes $N$ considered. We
would naturally conclude that the overall computational cost $ T \sim T_2$ is
cubic. In the opposite case where a very inefficient linear-scaling algorithm is employed for
the first part, $T \sim T_1 \gg T_2$, we would conclude to a linear-scaling type
behavior.
Of course, mathematically speaking such a way of reasoning is not correct 
since scaling laws are only meaningful in the {\it asymptotic} regime where $N$ goes
to infinity.  However, in practice only a {\it finite range of sizes is
considered} (here, between 2 and about 2000 active electrons) and it is important to be 
very cautious with the notion of scaling laws.
A more correct point of view consists in looking at the global performance of the
code in terms of total CPU time for a given range of system sizes, a given
compiler, and a given type of CPU core.

Finally, a last aspect concerns the memory footprint of the code whose
minimization turns out to be very advantageous. Indeed, the current trend in
supercomputer design is to increase the number of cores more rapidly than the
available total memory. As the amount of memory per core will continue to
decrease, it is very likely that programs will need to have a low memory
footprint to take advantage of exascale computers.
Another point is that when less memory is used less
electrical power is needed to perform the calculation: data movement from the
memory modules to the cores needs more electrical power than performing floating
point operations. Although at present time the power consumption is not yet a concern to software
developers, it is a key aspect in present design of the exascale machines to come.

In this section, the results discussed will be systematically presented by
using two different generations of Intel Xeon processors.
The first processor, referred to as {\em Core2}, is an Intel Xeon 5140, Core2
2.33~GHz, Dual core, 4~MiB shared L2 cache. The second one, referred to as {\em Sandy
Bridge}, is an Intel Xeon E3-1240 at 3.30~GHz, Quad core, 256~KiB L2
cache/core, 8~MiB shared L3 cache (3.4~GHz with turbo).
Note also that the parallel scaling of QMC being close to ideal (see next section), single-core
optimization is very interesting: the gain in execution time obtained on the
single-core executable is directly transferred to the whole parallel simulation.

\subsection{Improving the innermost expensive floating-point operations}

For the Core2 architecture, the practical algorithm presented above may be
further improved by first using the {\em unroll and jam} technique,\cite{unroll}
which consists in unrolling the outer loop and merging multiple outer-loop
iterations in the inner loop~:
\begin{verbatim}
do i=1, Number of electrons
 do k=1, Number of non-zero AOs for electron i, 2
  do j=1, Number of molecular orbitals
   C1(j,i) += A(j,indices(k  ,i))*B1(k  ,i) + &
              A(j,indices(k+1,i))*B1(k+1,i)
   C2(j,i) += A(j,indices(k  ,i))*B2(k  ,i) + &
              A(j,indices(k+1,i))*B2(k+1,i)
   ...
  end do
 end do
end do
\end{verbatim}
%est-ce bien connu des "computational chemists" ? Si non, ajouter quelque
%chose du genre '(use of memory to simulate extra missing registers)'
To avoid register spilling, the inner loop is split in two loops~:
one loop computing ${\bf C}_1,{\bf C}_2,{\bf C}_3$ and a second loop computing
${\bf C}_4,{\bf C}_5$. The load/store ratio is improved from 6/5 to 5/3 and
4/2

For the Sandy Bridge architecture, the external body is unrolled four times
instead of two, and the most internal loop is split in three loops: one loop
computing ${\bf C}_1,{\bf C}_2$, a second loop computing ${\bf C}_3,{\bf C}_4$, and a third loop
computing ${\bf C}_5$.
The load/store ratio is improved from 6/5 to 6/2 and 5/1.

%attention : ne pas aligner les accès mémoires n'empêche pas, sur x86, de vectoriser.
%Par contre, aligner les accès mémoire permet d'en réduire le coût et donc d'
%améliorer l'efficacité du code vectorisé. Il faut peut-être réécrire une phrase ou
%deux en conséquence
Then, all arrays were 256-bit aligned using compiler directives and
the first dimensions of all arrays were set to a multiple
of 8 elements (if necessary, padded with zeros at the end of each column) in
order to force a 256-bit alignment of every column of the matrices. These
modifications allowed the compiler to use {\em only} vector instructions to
perform the matrix products, both with the Streaming SIMD Extension (SSE)
or the Advanced Vector Extension (AVX) instruction sets.
The x86\_64 version of the MAQAO framework\cite{maqao} indicates that, as the
compiler unrolled twice the third loop (${\bf C}_5$), these three loops perform 16
floating point operations per cycle, which is the peak performance on this
architecture.

Finally, to improve the cache hit probability, blocking was used on the first
dimension of ${\bf B}_n$ (loop over $k$). In each block, the electrons (columns of
${\bf B}$) are sorted by ascending first element of the \texttt{indices} array in the block.
This increases the probability that columns of ${\bf A}$ will be in the cache for the
computation of the values associated with the next electron.

\begin{table*}
\hfill{}
\begin{tabular}{|l|ccc|ccc|}
\hline
                  & \multicolumn{3}{|c|}{\textbf{Core2}}           &   \multicolumn{3}{|c|}{\textbf{Sandy Bridge}}            \\
                  &  \textbf{Products} & \textbf{Inversion} & \textbf{Overall} &  \textbf{Products} & \textbf{Inversion} & \textbf{Overall} \\
\hline
Linpack (DP)      &               & 7.9 (84.9\%) &     &               & 24.3 (92.0\%) &        \\
\hline                                      
Peak              & 18.6          & 9.3          &     & 52.8          & 26.4          &      \\
Smallest system   &  9.8 (52.7\%) & 2.6 (28.0\%) & 3.3 & 26.6 (50.3\%) & 8.8  (33.3\%) & 6.3  \\
$\beta$-Strand    &  9.7 (52.2\%) & 4.3 (46.2\%) & 3.7 & 33.1 (62.7\%) & 13.7 (51.2\%) & 13.0 \\
$\beta$-Strand TZ &  9.9 (53.2\%) & 4.3 (46.2\%) & 4.5 & 33.6 (63.6\%) & 13.7 (51.2\%) & 14.0 \\
1ZE7              &  9.3 (50.0\%) & 5.2 (55.9\%) & 4.6 & 30.6 (57.9\%) & 15.2 (57.6\%) & 17.9 \\
1AMB              &  9.2 (49.5\%) & 5.6 (60.2\%) & 5.0 & 28.2 (53.4\%) & 16.2 (61.4\%) & 17.8 \\
\hline
\end{tabular}
\hfill{}
\caption{Single core performance (GFlops/s) of the matrix products (single precision),
inversion (double precision) and overall performance of QMC=Chem (mixed single/double
precision). The percentage of the peak performance is given in parentheses.
Core2 : Intel Xeon 5140, Core2 2.33GHz, Dual core, 4MiB shared L2 cache.
Sandy Bridge : Intel Xeon E3-1240, Sandy Bridge 3.30GHz, Quad core, 256KiB L2 cache/core, 8MiB shared L3 cache (3.4GHz with turbo).}
\label{tab2}
\end{table*}

The results obtained using the Intel Fortran Compiler XE 2011 are
presented in table~\ref{tab2} for both the Core2 and the Sandy Bridge
architectures. The single-core double-precision Linpack benchmark is also
mentioned for comparison.
The results show that the full performance of the matrix products is
already reached for the smallest system.  However, as opposed to dense matrix
product routines, we could not approach further the peak performance of
the processor since the number of memory accesses scales as the number of
floating point operations (both ${\cal O}(N^2)$): the limiting factor is
inevitably the data access.
Nevertheless, the DECAN tool\cite{decan} revealed that data access only adds a
30\% penalty on the pure arithmetic time, indicating an excellent use of the
hierarchical memory and the prefetchers.

\subsection{Single-core performance}

\subsubsection{Computational cost as a function of the system size} 

\begin{figure}
\begin{center}
\includegraphics[width=\columnwidth,keepaspectratio=true]{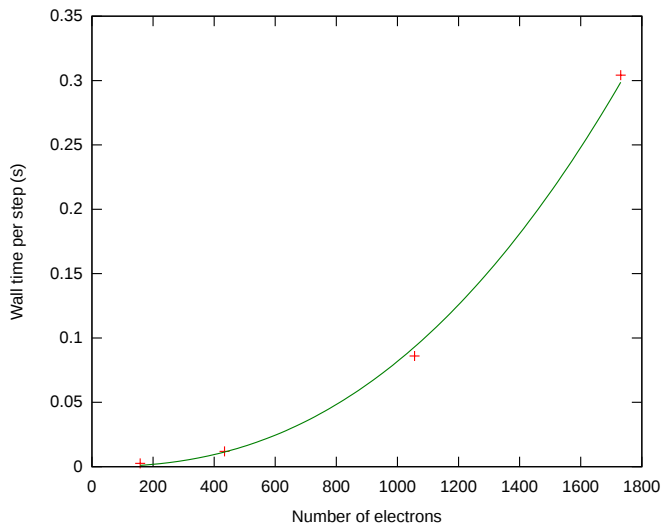}
\end{center}
\caption{Single-core scaling with system size.}
\label{fig2}
\end{figure}

\begin{table*}
\hfill{}
\begin{tabular}{|l|ccccc|}
\hline
            & \textbf{Smallest system} & \textbf{$\beta$-Strand} & \textbf{$\beta$-Strand TZ} & \textbf{1ZE7} & \textbf{1AMB} \\
\hline
RAM (MiB)   &  9.8            &  31            &  65               & 133      & 313     \\
\hline
\textbf{Core2}       &                 &                &                   &          &         \\
QMC step(s) &  0.0062         &  0.0391        &  0.0524           & 0.2723   & 0.9703  \\
Inversion   &  15\%           &  31\%          &  21\%             & 47\%     & 58\%    \\
Products    &  25\%           &  23\%          &  35\%             & 21\%     & 18\%    \\
\hline
\textbf{Sandy Bridge}&                 &                &                   &          &         \\
QMC step(s) &  0.0026         &  0.0119        &  0.0187           & 0.0860   & 0.3042  \\
Inversion   &  12\%           &  26\%          &  17\%             & 42\%     & 52\%    \\
Products    &  24\%           &  22\%          &  32\%             & 21\%     & 20\%    \\
\hline
\end{tabular}
\hfill{}
\caption{Single-core memory consumption and elapsed time for one VMC step. Values in \%
represent the percentage of the total CPU time.
Core2 : Intel Xeon 5140, Core2 2.33GHz, Dual core, 4MiB shared L2 cache.
Sandy Bridge : Intel Xeon E3-1240, Sandy Bridge 3.30GHz, Quad core, 256KiB L2 cache/core, 8MiB shared L3 cache (3.4GHz with turbo).}
\label{tab3}
\end{table*}

In table~\ref{tab3} the memory required together with the CPU time obtained for
one Monte Carlo (VMC) step for the five systems are presented using both
processors.
The two expensive computational parts (matrix products and inversion) are
distinguished.  A first remark is that the trends for both processors are very
similar so we do not need to make a distinction at this moment. A second remark
is that the memory footprint of QMC=Chem is particularly low. For the biggest
size considered (1731 electrons) the amount of RAM needed is only 313~MiB.
Finally, another important remark is that at small number of electrons the 
multiplicative part is dominant while this is not the case at larger sizes.
Here, the change of regime is observed somewhere between 400 and 1000
electrons but its precise location depends strongly on the number of
basis functions used. For example, for systems 3 and 4 corresponding to the
same molecule with a different number of basis functions, the multiplicative
part is still dominant for the larger basis set ($\beta$-strand with cc-pVTZ)
while it is no longer true for the smaller basis set ($\beta$-strand with
6-31G$^*$). In figure \ref{fig2} a plot
of the total computational time for the Sandy Bridge core as a function of the
number of electrons is presented. A standard fit of the curve with a polynomial 
form $N^{\gamma}$ leads to a $\gamma$-value of about 2.5. However, as
discussed above such a power is not really meaningful. From the data of
table~\ref{tab3} it is easy to extract the pure contribution related to the
inversion and a factor very close to 3 is obtained, thus illustrating that
for this linear algebra part we are in the asymptotic regime. For the
multiplicative part, the pure $N^2$ behavior is not yet recovered and we are in
an intermediate regime.  Putting together these two situations leads to some
intermediate scaling around 2.5.

\subsubsection{Sparsity} 
In our practical algorithm, for the matrix products we have chosen to
consider the ${\bf B}$ matrices as sparse as opposed to the ${\bf A}$ matrix which is
considered dense. The reason for that is that considering the matrix ${\bf A}$ sparse
would not allow us to write a stride-one inner loop. In
single precision, SSE instructions executed on Intel processors can perform up
to 8 instructions per CPU cycle (one 4-element vector ADD instruction and one
4-element vector MUL instruction in parallel). Using the latest AVX instruction
set available on the Sandy Bridge architecture, the width of the SIMD vector
registers have been doubled and the CPU can now perform up to 16 floating point
operations per cycle. A necessary condition for enabling vectorization is a
stride-one access to the data. This implies that using a sparse representation
of ${\bf A}$ would disable vectorization, and reduce the maximum number of floating
operations per cycle by a factor of 4 using SSE (respectively 8 using AVX). If
matrix ${\bf A}$ has more than 25\% (respectively 12.5\%) non-zero elements, using
a sparse representation is clearly not the best choice. This last result is a
nice illustration of the idea that the efficiency of the formal mathematical
algorithm depends on the core architecture.

\subsubsection{Inversion step} 
Now, let us consider the inversion step which is the dominant CPU-part for the
big enough systems (here, for about a thousand electrons and more).  In
Table~\ref{tab2} the performance in GFlops/s of the inversion step is presented
for both processors. For comparisons the theoretical single-core peak and
single-core Linpack performance are given. The first column gives the overall
performance of the code while the second column is specific to the inversion
part. As seen the performance of both parts increases with the number of
electrons. For largest systems the performance represents more than 50\% of
the peak performance of each processor. For the largest system the whole code
has a performance of about 54\% of the peak performance for the Core2, and
about 61\%  for the Sandy Bridge. The performance is still better for the
inversion part: 60\% for the Core2 and 67\% for the Sandy Bridge.

\subsubsection{Determinant calculation compared to spline interpolation}

Most authors use three-dimensional spline representations of the molecular
orbitals in order to compute in constant time the values, first derivatives and
Laplacians of one electron in one molecular orbital, independently of the size
of the atomic basis set. This approach seems efficient at first sight, but the
major drawback is that the memory required for a single processor can become
rapidly prohibitive since each molecular orbital has to be pre-computed on a
three-dimensional grid.  To overcome the large-memory problem, these authors
use shared memory approaches on the computing nodes, which implies coupling
between the different CPU cores. In this paragraph, we compare the wall time
needed for spline interpolation or computation of the values, first derivatives
and Laplacians of the wave function at all electron positions.

Version 0.9.2 of the Einspline package\cite{einspline} was used as a reference
to compute the interpolated values, gradients and Laplacians of 128 MOs
represented on $23\times21\times29$ single precision arrays. The ``multiple
uniform splines'' set of routines were used.  To evaluate the value, gradient
and Laplacian of one molecular orbital at one electron coordinate,
an average of 1200 CPU cycles was measured using LIKWID\cite{likwid} on the Core2 processor versus 850
CPU cycles on the Sandy Bridge processor. Even if the interpolation is done
using a very small amount of data and of floating point operations, it is bound by
the memory latency. Indeed, the needed data is very unlikely to be in the CPU
cache and this explains why the number of cycles per matrix element is quite
large.
As our code uses a very small amount of memory, and as the computationally
intensive routines are very well vectorized by the compiler, the computation
of the matrix elements is bound by the floating point throughput
of the processor.

\begin{table*}
\hfill{}
\begin{tabular}{|l|ccc|ccc|}
\hline
          & \multicolumn{3}{|c|}{\textbf{Core2}}           &   \multicolumn{3}{|c|}{\textbf{Sandy Bridge}}            \\
          &  \textbf{QMC=Chem} & \textbf{Einspline} & \textbf{ratio} & \textbf{QMC=Chem} & \textbf{Einspline} & \textbf{ratio} \\
\hline
\hline
Smallest system  &   16.7 &   15.0 & 1.11 &    9.2  &    10.6   &  0.87      \\
$\beta$-Strand TZ&  177.3 &  113.0 & 1.57 &   81.7  &    80.1   &  1.02      \\
1ZE7             &  783.5 &  669.1 & 1.17 &  352.0  &   473.9   &  0.74      \\
1AMB             & 2603.0 & 1797.8 & 1.45 & 1183.9  &  1273.5   &  0.93      \\
\hline
\end{tabular}
\hfill{}
\caption{Number of million CPU cycles needed for the computation of the values, gradients and Laplacians
of the molecular orbitals using the Einspline package and using our implementation for the Core2
and the Sandy Bridge micro-architectures. The ratio QMC=Chem / Einspline is also given.}
\label{tab5}
\end{table*}

The number of cycles needed to build the ${\bf C}_1 \dots {\bf C}_5$ matrices
is the number of cycles needed for one matrix element scaled by
the number of matrix elements $N_{\alpha}^2+N_{\beta}^2$.
Table~\ref{tab5} shows the number of CPU cycles needed to build
the full ${\bf C}_1 \dots {\bf C}_5$ matrices for a new set of electron positions
using spline interpolation or using computation. The computation includes
the computation of the values, gradients and Laplacians of the atomic orbitals
(matrices ${\bf B}_1 \dots {\bf B}_5$) followed by the matrix products.

Using a rather small basis set (6-31G$^*$), the computation of the matrices in the 
158-electron system is only 10\% slower than the interpolation on the Core2
architecture.  Using a larger basis set (cc-pVTZ), the computation is only 57\%
slower.

As the frequency is higher in our Sandy Bridge processor than in our Core2
processor, we would have expected the number of cycles of one memory latency to
increase, and therefore we would have expected the Einspline package to be less
efficient on that specific processor. One can remark that the memory latencies
have been dramatically improved from the Core2 to the Sandy Bridge architectures
and the number of cycles for the interpolation decreases.

The full computation of the matrix elements benefits from the improvement in
the memory accesses, but also from the enlargement of the vector registers from
128 bits to 256 bits.  This higher vectorization considerably reduces the
number of cycles needed to perform the calculation such that in the worst case
(the largest basis set), the full computation of the matrix elements takes as
much time as the interpolation. In all other cases, the computation is faster
than the spline interpolation.
Finally, let us mention that as the memory controller is directly attached to
the CPU, on multi-socket computing nodes the memory latencies are higher
when accessing a memory module attached to another CPU (NUMA architecture). 

\section{Parallelism: implementing a universal, dynamic, and fault-tolerant scheme}
\label{parallel}

Our objective was to design a program that could take maximum advantage of
heterogeneous clusters, grid environments, the petaflops platforms available
now and those to come soon (exascale). 

To achieve the best possible parallel speed-up on any hardware, all the
parallel tasks have to be completely decoupled. Feldman {\em et al} have shown
that a naive implementation of parallelism does not scale well on commodity
hardware.\cite{goddard} Such bad scalings are also expected to be observed on
very large scale simulations. Therefore, we chose an implementation where each
CPU core realizes a QMC run with its own population of walkers independently of
all the other CPU cores. The run is divided in {\em blocks} over which the
averages of the quantities of interest are computed. The only mandatory
communications are the {\em one-to-all} communication of the input data and the
{\em all-to-one} communications of the results, each result being the Monte
Carlo average computed with a single-core executable. If a single-core
executable is able to
start as soon as the input data is available and stop at any time sending an
average over all the computed Monte Carlo steps, the best possible parallel
speed-up on the machine can always be obtained. This aspect is detailed in
this section.

\subsection{Fault-tolerance}
Fault-tolerance is a critical aspect since the mean time before failure
(MTBF) increases with the number of hardware components~: using $N$ identical
computing nodes for a singe run multiplies by $N$ the probability of failure of
the run. If one computing node is expected to fail once a year, a run using 365
computing nodes is not expected to last more than a day. As our goal is the use
both of massive resources and commodity clusters found in laboratories,
hardware failure is at the center of our software design.

The traditional choice for the implementation of parallelism is the use of the
Message Passing Interface (MPI).\cite{mpi} Efficient libraries are proposed on
every parallel machine, and it is probably the best choice in most situations.
However, all the complex features of MPI are not needed for our QMC
program, and it does not really fit our needs: in the usual MPI
implementations, the whole run is killed when one parallel task is known not be
able to reach the {\tt MPI\_Finalize} statement. This situation occurs when a
parallel task is killed, often due to a system failure (I/O error, frozen
computing node, hardware failure, etc). For deterministic calculations where the
result of every parallel task is required, this mechanism prevents from
unexpected dead locks by immediately stopping a calculation that will never
end.
In our implementation, as the result of the calculation of a block is a
Gaussian distributed random variable, removing the result of a block from the
simulation is not a problem since doing that does not introduce any bias in the
final result. Therefore, if one computing node fails, the rest of the simulation
should survive.

\begin{figure}
\begin{center}
\includegraphics[width=\columnwidth,keepaspectratio=true]{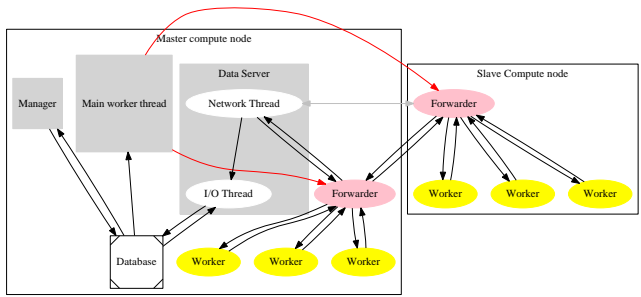}
\end{center}
\caption{Overview of the QMC=Chem architecture.}
\label{fig3}
\end{figure}

We wrote a simple Python TCP client/server application to handle parallelism.
To artificially improve the bandwidth, all network transfers are compressed
using the Zlib library,\cite{zlib} and the results are transferred
asynchronously in large packets containing a collection of small messages.
Similarly, the storage of the results is executed using a non-blocking 
mechanism.
The computationally intensive parts were written using the IRPF90 code
generator,\cite{irpf90} to produce efficient Fortran code that is also easy to
maintain.
The architecture of the whole program is displayed in figure \ref{fig3}.

\subsection{Program interface}
Our choice concerning the interaction of the user with the program was not to
use the usual ``input file and output file'' structure. Instead, we chose to
use a database containing all the input data and control parameters of the
simulation, and also the results computed by different runs. A few simple
scripts allow the interaction of the user with the database.  This choice has
several advantages:
\begin{itemize}
\item The input and output data are tightly linked together. It is always
possible to find to which input corresponds output data.
\item If an output file is needed, it can be generated on demand using
different levels of verbosity
\item Graphical and web interfaces can be trivially connected to the program 
\item Simple scripts can be written by the users to manipulate the computed
data in a way suiting their needs.
\end{itemize}
Instead of storing the running average as the output of a run, we store all the
independent block-averages in the database, and the running averages are
post-processed on demand by database queries. There are multiple benefits from
this choice:
\begin{itemize}
\item Checkpoint/restart is always available
\item It is possible to compute correlations, combine different random variables,
{\em etc}, even when the QMC run is finished
\item Combining results computed on different clusters consists in simply
merging the two databases, which allows automatically the use of the program on
computing grids\cite{qmcgrid}
\item Multiple independent jobs running on the same cluster can read/write in
the same database to communicate via the file system. This allows to gather
more and more resources as they become available on a cluster, or to run a
massive number of tasks in a {\em best effort} mode.\cite{besteffort}
 \end{itemize}

\subsection{Error-checking}
We define the critical data of a simulation as the input data that
characterizes uniquely a given simulation. For instance, the molecular
coordinates, the molecular orbitals, the Jastrow factor parameters are critical
data since they are fixed parameters of the wave function during a QMC run. In 
contrast, the number of walkers of a simulation is not critical data for a
VMC run since the results of two VMC simulations with a different number of
walkers can be combined together.
A 32-bit cyclic redundancy code (CRC-32 key) is associated with the critical
data to characterize a simulation. This key will be used to guarantee that the
results obtained in one
simulation will never be mixed with the results coming from another simulation
and corrupt the database. It will also be used to check that the input data has
been well transferred on every computing node.

\subsection{Program execution}
When the program starts its execution, the {\em manager} process runs on the master
node and spawns two other processes: a {\em data server} and a {\em main worker
process}.

At any time, new clients can connect to the
data server to add dynamically more computational resources to a running
calculation, and some running clients can be terminated without stopping the whole
calculation.
The manager periodically queries the database and computes the running 
averages using all the blocks stored in the database. It controls the
running/stopping state of the workers by checking if the stopping condition is
reached (based for example on the wall-clock time, on the error bar of the
average energy, a Unix signal, etc).

When running on super-computers, the main worker process spawns one single
instance of a {\em forwarder} on each computing node given by the batch scheduler
system using an MPI launcher. As soon as the forwarders are started the MPI
launcher terminates, and each forwarder connects to the data server to retrieve
the needed input data. The forwarder then starts multiple {\em workers} on the
node with different initial walker positions.

Each worker is an instance of the single-core Fortran executable, connected to
the forwarder by Unix pipes. Its behavior is the following~:
\begin{verbatim}
 while ( .True. )
 {
    compute_a_block_of_data();
    send_the_results_to_the_forwarder();
 }
\end{verbatim}
Unix signals \verb'SIGTERM' and \verb'SIGUSR2' are trapped to trigger the
\verb'send_the_results_to_the_forwarder' procedure followed by the termination
of the process. Using this mechanism, any single-core executable can be stopped
{\em immediately} without losing a single Monte Carlo step. This aspect is
essential to obtain the best possible speed-up on massively parallel machines.
Indeed, using the matrix product presented in the previous section makes
the CPU time of a block non-constant. Without this mechanism the run would
finish when the last CPU finishes, and the parallel efficiency would be 
reduced when using a very large number of CPU cores.

\begin{figure}
\begin{center}
\includegraphics[width=\columnwidth,keepaspectratio=true]{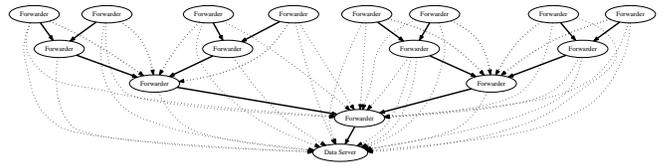}
\end{center}
\caption{Connections of the forwarders with the data server.}
\label{fig4}
\end{figure}

While the workers are computing the next block, the forwarder sends the
current results to the data-server using a path going through other forwarders.
The forwarders are organized in a binary tree as displayed in figure \ref{fig4}: every
node of the tree can send data to all its ancestors, in order to deal with
possible failures of computing nodes. This tree-organization reduces the number
of connections to the data server, and also enlarges the size of the messages
by combining in a single message the results of many forwarders.

At the end of each block, the last walker positions are sent from the worker to
the forwarder. The forwarder keeps a fixed-sized list of $N_{\rm kept}$ walkers enforcing the
distribution of local energies~: when a forwarder receives a set of $N$ walkers,
it appends the list of new walkers to its $N_{\rm kept}$ list, and sorts the $N_{\rm kept}+N$ list
by increasing local energies. A random number $\eta$ is drawn to keep all list entries
at indices $\lfloor \eta + i(N_{\rm kept}+N)/N_{\rm kept}\rfloor, i=\{1,\dots,N_{\rm kept}\}$
%'A random number is drawn to remove all list entries with an odd or even list
% index.' : pourquoi ? TODO
After a random timeout, if the forwarder is idle, it sends its list of walkers
to its parent in the binary tree which repeats the list merging process.
Finally, the data server receives a list of walkers, merges it with its own
list and writes it to disk when idle. This mechanism ensures that the walkers
saved to disk will represent homogeneously the whole run and avoids sending all
the walkers to the data server. These walkers will be used as new starting
points for the next QMC run.

Using such a design the program is robust to system failures. Any computing node
can fail with a minimal impact on the simulation:
\begin{itemize}
\item If a worker process fails, only the block being computed by
this worker is lost. It does not affect the forwarder to which it is linked.
\item If a forwarder fails, then only one computing node is lost thanks to the
redundancy introduced in the binary tree of forwarders. 
\item The program execution survives short network disruption (a fixed timeout
parameter). The data will arrive to the data server when the network becomes
operational again.
\item The disks can crash on the computing nodes~: the temporary directory used
on the computing nodes is a RAM-disks (\verb'/dev/shm').
\item The shared file system can fail as the single-core static executable, the
python scripts and input files are broadcast to the RAM-disks of the compute
nodes with the MPI launcher when the run starts
\item Redundancy can be introduced on the data server by running multiple jobs
using the same database. Upon a failure of a data server, only the forwarders
connected to it will be lost.
\item In the case of a general power failure, all the calculations can be
restarted without losing what has already been stored in the database.
\end{itemize}
% RESULTATS PARALLELES
Finally, we have left the possibility of using different executables connected
to the same forwarder. This will allow a combined use of pure CPU executables
with hybrid CPU/GPU and CPU/MIC executables, to use efficiently all the available
hardware. The extension to hybrid architectures will be the object of a future work.

\subsection{Parallel speed-up}

\begin{table*}
\hfill{}
\begin{tabular}{|l|ccccc|}
\hline
                           & \textbf{Smallest system} &  \textbf{$\beta$-Strand} &  \textbf{$\beta$-Strand TZ} &  \textbf{1ZE7} & \textbf{1AMB}   \\
\hline                                                                                                                                              
$N$                        &       158                &  434                     & 434                         & 1056           & 1731             \\
$N_{\rm basis}$            &       404                &  963                     & 2934                        & 2370           & 3892             \\
\hline                                                                                                                                              
\% of non-zero$^a$         &       81.3\%             &  48.4\%                  & 73.4\%                      & 49.4\%         &37.1\%            \\
MO coefficients $a_{ij}$   &      (99.4\%)            &  (76.0\%)                & (81.9\%)                    &(72.0\%)        &(66.1\%)          \\
($A_{ij} \neq 0$)            &                          &                          &                             &                &                  \\
\hline                                                                                                                                              
Average \% of non-zero     &                          &                          &                             &                &                  \\
basis functions            
$\chi_i({\bf r}_j)$        &       36.2\%             &     14.8\%               & 8.2\%                       & 5.7\%          & 3.9\%            \\
(${B_1}_{ij} \neq 0$)        &                          &                          &                             &                &                  \\
\hline                                                                                                                                              
Average number of          &                          &                          &                             &                &                  \\
non-zero elements          &       146                &     142                  &  241                        &  135           &  152             \\
per column of ${B_1}_{ij}$ &                          &                          &                             &                &                  \\
\hline
\end{tabular}
\hfill{}
\caption{System sizes, percentage of non-zero molecular orbital coefficients
and average percentage of non-zero Atomic Orbital values.
$^a$ Zero MO coefficients are those below $10^{-5}$. These are given for localized
orbitals, and for canonical orbitals in parentheses.
$^b$ Zero AO matrix elements are those for which the radial component of
the basis function has a value below $10^{-8}$ for given electron positions.}
\label{tab1}
\end{table*}

The benchmarks presented in this section were performed on the Curie machine
(GENCI-TGCC-CEA, France). Each computing node is a dual socket Intel Xeon
E5-2680: 2$\times$(8 cores, 20~MiB shared L3-cache, 2.7~GHz) with 64~GiB of RAM.
The benchmark is a DMC calculation of the $\beta$-strand system with the cc-PVTZ
basis set (Table~\ref{tab1}) using 100 walkers per core performing 300 steps in each
block. Note that these blocks are very short compared to realistic simulations
where the typical number of steps would be larger than 1000 to avoid the
correlation between the block averages.

\subsubsection{Intra-node}
The CPU consumption of the forwarder is negligible (typically 1\% of the CPU
time spent in the single-core executables). The speed-up with respect to the number
of sockets is ideal. Indeed the single-core binaries do not communicate between each
other, and as the memory consumption per core is very low, each socket never
uses memory modules attached to another socket. When multiple cores on the
same socket are used we observed a slow-down for each core due to the sharing
of the L3-cache and memory modules. Running simultaneously 16 instances of the
single-core binaries on our benchmark machine yields an increase of 10.7\% of the
wall-clock time compared to running only one instance. For a 16-core run, we obtained
a 14.4$\times$ speed-up (the Turbo feature of the processors was de-activated for this
benchmark).

\subsubsection{Inter-node}

In this section the wall-clock time is measured from the very beginning to the
very end of the program execution using the standard {\em
GNU time} tool. Hence, the wall-clock time includes the initialization and
finalization steps.

The initialization step includes
\begin{itemize}
\item Input file consistency checking
\item Creating a gzipped tar file containing the input files (wave function
parameters, simulation parameters, a pool of initial walkers), the Python scripts
and static single-core executable needed for the program execution on the slave
nodes
\item MPI Initialization
\item Broadcasting the gzipped tar file via MPI to all the slave nodes
\item Extracting the tar file to the RAM-disk of the slave nodes
\item Starting the forwarders
\item Starting the single-core instances
\end{itemize}
Note that as no synchronization is needed between the nodes, the computation
starts as soon as possible on each node.

The finalization step occurs as follows. When the data server receives a
termination signal, it sends a termination signal to all the forwarders that
are leaves in the tree of forwarders. When a forwarder receives such a signal,
it sends a {\tt SIGTERM} signal to all the single-core binary instances of the
computing node which terminate after sending to the forwarder the averages
computed over the truncated block. Then, the forwarder sends this data to its
parent in the binary tree with a termination signal, and sends a message to the
data server to inform it that it is terminated. This termination step walks
recursively through the tree. When all forwarders are done, the data server
exits.
Note that if a failure happened on a node during the run, the data server never
receives the message corresponding to a termination of the corresponding
forwarder. Therefore when the data server receives the termination signal
coming from the forwarders tree, if the data server is still running after a
given timeout it exits.

\begin{figure}
\begin{center}
\includegraphics[width=\columnwidth,keepaspectratio=true]{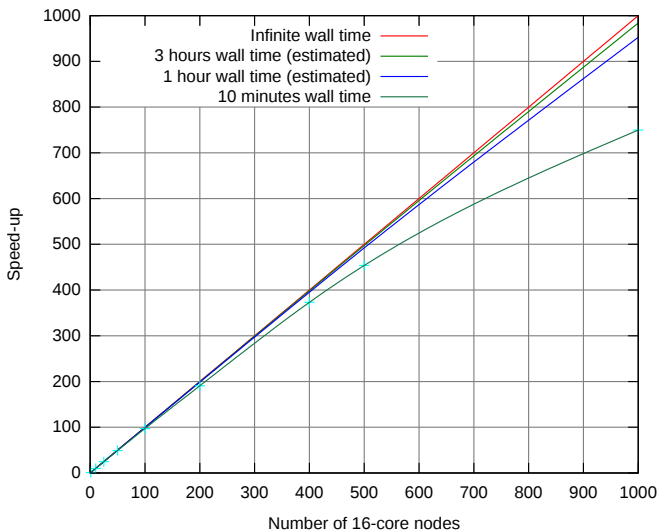}
\end{center}
\caption{Parallel speed-up of QMC=Chem with respect to 16-core compute nodes (Reference
is one 16-core node).}
\label{fig5}
\end{figure}

\begin{table*}
\hfill{}
\small
\begin{tabular}{|l|ccc|ccc|c|}
\hline
\textbf{Number of} &  \multicolumn{3}{|c|}{\textbf{10 minutes}}  & \multicolumn{3}{|c|}{\textbf{60 minutes}}     & \textbf{180 minutes} \\
\textbf{16-core}   &                                      &&&      \multicolumn{3}{|c|}{\textbf{(estimated)}}& \textbf{(estimated)} \\
\textbf{Nodes}     &  \textbf{CPU (s)} &  \textbf{Wall (s)} & \textbf{Speed-up} & \textbf{CPU(s)} &  \textbf{Wall(s)} & \textbf{Speed-up} & \textbf{Speed-up} \\
\hline
      1 &     9627 & 625 &   1.0 &       57147 &     3625 &      1.00 &    1.00 \\
        &          &     &       &  {\em 57332}&{\em 3629}& {\em 1.00}&         \\
     10 &    95721 & 627 &   9.9 &      570921 &     3627 &      9.98 &    9.99 \\
     25 &   239628 & 629 &  24.7 &     1427628 &     3629 &     24.95 &   24.98 \\
     50 &   477295 & 631 &  49.1 &     2853295 &     3631 &     49.85 &   49.95 \\
    100 &   952388 & 636 &  97.2 &     5704388 &     3636 &     99.52 &   99.84 \\
        &          &     &       &{\em 5708422}&{\em 3638}&{\em 99.32}&         \\
    200 &  1869182 & 637 & 190.5 &    11373182 &     3637 &    198.36 &  199.45 \\
    400 &  3725538 & 648 & 373.3 &    22733538 &     3648 &    395.30 &  398.42 \\
    500 &  4479367 & 641 & 453.7 &    28239367 &     3641 &    491.98 &  497.31 \\
   1000 &  8233981 & 713 & 749.7 &    55753981 &     3713 &    952.50 &  983.86 \\
\hline
\end{tabular}
\hfill{}
\caption{Data relative to the scaling curve (Figure~\ref{fig5}).
CPU time is the cumulated CPU time spent only in the Fortran executables,
Wall time is the measured wall-clock time, including initialization and
finalization steps (serial).
The 10-minutes run were measured, and the longer runs are estimated from the
10-minutes run data.  Two checks were measured for the 60-minutes runs with 1
and 100 nodes (in italics).}
\label{tab6}
\end{table*}

We prepared a 10-minutes run for this section to compute the parallel
speed-up curve as a function of the number of 16-core nodes given in
Figure~\ref{fig5}. The data corresponding to this curve are given in Table~\ref{tab6}.
The reference for the speed-up is the one-node run. The
speed-up for $N$ nodes is computed as:
\begin{equation}
\frac{t_{\rm CPU}(N) / t_{\rm Wall}(N)}{t_{\rm CPU}(1) / t_{\rm Wall}(1)}
\end{equation}
The initialization time was 9 seconds for the single node run, and 22 seconds
for the 1000 nodes run.
The finalization time was 13 seconds for the single node run, and 100 seconds
for the 1000 nodes run.

Apart from the initialization and finalization steps (which obviously do not
depend on the total execution time), the parallel speed-up is ideal.  This
allowed us to estimate the speed-ups we would have obtained for a 1-hour run
and for a 3-hours run. For instance, to estimate the one-hour run we added 50
minutes to the wall-clock time and 50 minutes$\times$16$\times$number of nodes
$\times$0.99 to the CPU time. The 99\% factor takes account of the CPU
consumption of the forwarder for communications. Our simple model was checked
by performing a one-hour run on one node, and a one-hour run on 100 nodes. An
excellent agreement with the prediction was found: a 99.5$\times$ speed-up was
predicted for 100 nodes and a 99.3$\times$ speed-up was measured.

Finally, a production run was made using 76,800 cores of Curie (4,800~nodes) on
the $\beta$-strand molecule with a cc-pVTZ basis set via 12 runs of 40 nodes,
and a sustained performance of 960~TFlops/s was measured. All the details and
scientific results of this application will be presented elsewhere.\cite{alz}

\section{Summary}
\label{summary}

Let us summarize the main results of this work. First, to enhance the computational efficiency of the expensive 
innermost floating-point operations (calculation and multiplication of
matrices) we propose to take advantage of the highly localized character of the {\it atomic} Gaussian basis functions, in contrast
with the standard approaches using localized {\it molecular} orbitals.
The advantages of relying on atomic localization have been illustrated on a series of molecules of increasing sizes
(number of electrons ranging from 158 to 1731). In this work, it is emphasized that the notion of scaling of the computational 
cost as a function of the system size has to be considered with caution.
Here, although the algorithm proposed is formally quadratic it displays a small enough prefactor to become very efficient in the
range of number of electrons considered. Furthermore, our implementation of the linear-algebra computational part has allowed to enlighten a
fundamental issue rarely discussed, namely the importance of taking into consideration the close links between algorithmic structure
and CPU core architecture. Using efficient techniques and optimization tools for enhancing single-core performance, this point has been
illustrated in various situations. Remark that this aspect is particularly important: as the parallel speed-up is very good,
the gain in execution time obtained for the single-core executable will also be effective in the total parallel simulation.

In our implementation we have chosen to minimize the memory
footprint. This choice is justified first by the fact that today the amount of memory per
CPU core tends to decrease and second by the fact that small memory footprints
allow in general a more efficient usage of caches. In this spirit, we propose
not to use 3D-spline representation of the molecular orbitals as usually done. We have shown that this can be realized without 
increasing the CPU cost.
For our largest system with 1731 electrons, only 313~MiB of memory
per core was required. As a consequence the key limiting factor of our code is only the available CPU time and neither the
memory nor disk space requirements, nor the network performance. Let us re-emphasize that this feature is well aligned with the current trends
in computer architecture for large HPC systems.

Finally, let us conclude by the fact that there is no fundamental reason why
the implementation of such a QMC simulation environment which has been validated at petaflops level could
not be extended to exascale.

\section*{Acknowledgments}
AS and MC would like to thank ANR for support under Grant No ANR 2011 BS08 004 01.
This work was possible thanks to the generous computational support from CALMIP
(Universit\'e de Toulouse), under the allocation 2011-0510, GENCI under the
allocation GEN1738, CCRT (CEA), and PRACE under the allocation RA0824.
The authors would also like to thank Bull, GENCI and CEA for their help in
this project.

\bibliography{p}

\begin{thebibliography}{39}
\bibitem[{bes()}]{besteffort}
\bibinfo{note}{When cluster resources are unused, a QMC job starts. When
  another user requests the resources, the QMC job is killed.}
\end{thebibliography}

\end{document}